\documentclass{elsart}
\usepackage{epsfig}
\usepackage{lineno}

\begin{document}
\runauthor{Marco Battaglia}
\begin{frontmatter}
\title{Studies of Vertex Tracking with SOI Pixel Sensors for Future Lepton Colliders}
\author[UCSC,LBNL,CERN]{Marco Battaglia,}
\author[LBNL]{Devis Contarato,}
\author[LBNL]{Peter Denes,}
\author[HEPHY]{Dietrich Liko,}
\author[INFN]{Serena Mattiazzo,}
\author[INFN]{Devis Pantano}
\address[UCSC]{Santa Cruz Institute of Particle Physics, 
University of California at Santa Cruz, CA 95064, USA}
\address[LBNL]{Lawrence Berkeley National Laboratory, 
Berkeley, CA 94720, USA}
\address[CERN]{CERN, CH-1211 Geneva 23, Switzerland}
\address[HEPHY]{Institut f\"ur Hochenergiephysik, A-1050 Wien, Austria}
\address[INFN]{Dipartimento di Fisica, Universit\'a di Padova and INFN,
Sezione di Padova, I-35131 Padova, Italy}

\begin{abstract}
This paper presents a study of vertex tracking with a beam hodoscope consisting of three layers of 
monolithic pixel sensors in SOI technology on high-resistivity substrate. We study the track extrapolation 
accuracy, two-track separation and vertex reconstruction accuracy in $\pi^-$~Cu interactions with 150 and 
300~GeV/$c$ pions at the CERN SPS. Results are discussed in the context of vertex tracking at future lepton 
colliders.
\end{abstract}
\end{frontmatter}

\typeout{SET RUN AUTHOR to \@runauthor}

%\linenumbers

\section{Introduction}

Vertex tracking is one of the capabilities most crucial to a detector at future lepton colliders and flavour factories.
Much of the anticipated physics program at a high energy lepton collider, such as the ILC, CLIC or a muon collider 
relies on the ability of efficiently discriminating heavy ($t$, $b$, $c$ and $\tau$) from light 
fermions~\cite{Hansen:2003sb,Battaglia:2010ce}. 
If the Higgs boson exists at a mass of $\simeq$ 125~GeV, as possibly indicated by the recent preliminary 
LHC~\cite{ATLAS,cms} and Tevatron~\cite{tevnph} results, the precise determination of its couplings to fermions will be 
essential to test whether the Brout-Englert-Higgs mechanism of electro-weak symmetry breaking is also responsible for fermion 
mass generation. Experiments at lower energy facilities, such as the Super KEKB~\cite{Dolezal:2010zz} and SuperB~\cite{Bona:2007qt} 
$b$-factories, require high resolution vertex tracking for time-dependent measurements, which are crucial to the physics program 
in heavy flavour physics~\cite{Forti:2011zz}.
Because the momentum of particles from $B$ meson decays at the $\Upsilon$(4S) is low and the sensitivity to short-lived hadrons 
produced at high energy colliders must extend to most, if not all, of their charged decay products, excellent track 
extrapolation  accuracy at low momentum is of paramount importance. This motivates the development of thin Si sensors.
Monolithic pixel sensors, such as CMOS active pixel sensors and DEPFET sensors, have emerged from the R\&D effort 
for future $e^+e^-$ colliders as the best suited solution for high-resolution vertex tracking with a minimal 
material budget in a low to moderate radiation environment~\cite{Besson:2006dd,Marinas:2011zz,Rizzo:2011zz}. 
More recently, monolithic pixel sensors on high resistivity substrate have demonstrated superior performance owing to the 
larger amount of charge collected, ensuring high detection efficiency, and the faster collection time, reducing also 
the effect of radiation damage~\cite{Dorokhov:2011zz,Deveaux:2011zz}. 

Among the technologies offering a CMOS process over a high resistivity sensitive volume, Silicon-On-Insulator (SOI), 
pioneered a decade ago~\cite{Marczewski:2004zz}, has been extensively tested in the last six years, within a collaborative 
R\&D effort developed in partnership with KEK and Lapis Semiconductor Co.\, Ltd. (formerly OKI Semiconductor). With the 
mitigation of the back-gating effect using a buried $p$-well (BPW) to protect the in-pixel CMOS circuitry~\cite{soi2}, 
SOI pixel sensor prototypes have demonstrated high detection efficiency and single point resolution at the micron 
level~\cite{beam2010-soi}, also after thinning of the handle wafer~\cite{beam2011-soi}. 

In this paper we illustrate the vertex tracking performance of a hodoscope composed of SOI pixel sensors based on the 
study of inelastic interaction of high energy pions in a beam test (SOIPIX) at the CERN SPS in Fall 2011. 
We present the results obtained for tracking resolution, two-track separation and vertex resolution, compare our 
data to simulation and discuss the relation to requirements for future colliders.

\section{Experimental Setup and Data Analysis}

The SOIPIX beam test experiment has been carried out on the SPS beam-line H4 in the CERN North Area with a beam 
hodoscope made of three layers of ``SOImager-2'' prototype chips, designed at LBNL and produced in 0.2~$\mu$m Lapis 
technology (formerly OKI) on $n$-type SOI wafers with a resistivity of the handle wafer of $\simeq$~700~$\Omega\cdot$cm. 
The sensor sensitive area is a 3.5$\times$3.5~mm$^2$ matrix of 256$\times$256 pixels arrayed on a 13.75~$\mu$m pitch, 
read out through four parallel arrays of 64 columns each~\cite{beam2010-soi}. The sensor detection performances for high 
energy particles and soft X-rays have already been presented in details in~\cite{beam2010-soi,beam2011-soi,xray-soi}. 
In this setup, a single sensor, thinned to 75~$\mu$m (singlet), is located upstream from a Cu target to define the 
position of impact of the incoming pion. 
A pair of 260~$\mu$m-thick sensors spaced by 9.4~mm (doublet) reconstructs the trajectories of the charged particles 
emerging from the target. This spacing is comparable to that of the two layers of the PXD pixel detector in 
BELLE-II and to the distance between the first and second layers of a LC vertex tracker.
The 3~mm-thick Cu target is inserted between the detector singlet and the doublet, 15~mm upstream from the first 
layer of the SOI pixel doublet. The extrapolation distance from the first doublet layer and the target corresponds to the 
design distance between the interaction point and Layer0 at SuperB and the first layer of the vertex tracker of an $e^+e^-$ 
linear collider at $\sqrt{s}$ = 250 - 500~GeV, while for a multi-TeV $e^+e^-$ collider, such as CLIC, the beam induced background 
requires a stay-clear radius of $\sim$25-30~mm. 
The use of a thin sensor for the singlet minimises the interaction effects upstream from the target. Data have been collected 
with $\pi^-$s of 150 and 300~GeV/$c$ momentum. Runs with 300~GeV/$c$ pions have also been taken without the target, for 
alignment and calibration purposes. We operate the sensors at a depletion voltage, $V_d$, of 50~V, corresponding to a 
depleted thickness of $\sim$100~$\mu$m in the doublet sensors and $\sim$60~$\mu$m in the thin singlet sensor.

The data acquisition system consists of a custom analog board pigtailed to a commercial FPGA development board, equipped 
with a Xilinx Virtex-5 FPGA used as control board~\cite{Battaglia:2009daq}. The sensor analog outputs are fed to
100~MS/s 14-bit ADCs through independent analog differential inputs. Digitised data are formatted and transferred to 
the DAQ computer via a USB-2.0 link at a rate of 25~Mbytes/s. 
Measurements are performed with the chip clocked at 12.5~MHz, corresponding to an 80~ns read-out time per pixel. 
The readout is synchronised with the SPS extraction pulse, such that a sequence of 960 frames from each of the three sensors 
are read-out during the 9.6~s-long SPS spill. Detectors are cooled using forced airflow and the temperature near the 
sensor surface is in the range (28$\pm$1)$^{\circ}$C during data taking. Data sparsification and zero suppression is 
performed on-line using a custom {\tt Root}-based~\cite{Brun:1997pa} program. Sensors are scanned for seed pixels with 
signal exceeding a preset threshold in noise units. For each seed, the 7$\times$7 pixel matrix centred around the seed 
position is selected and stored on file. The pixel pedestal and noise values are updated at the end of each SPS spill, 
in order to follow possible drifts of their baselines throughout the data taking. Data are stored in {\tt Root} format and 
subsequently converted into {\tt lcio} format~\cite{Gaede:2005zz} for offline analysis. 

The data analysis is based on a set of custom processors in the {\tt Marlin} reconstruction framework~\cite{Gaede:2006pj} 
to perform cluster centre-of-gravity reconstruction, pattern recognition, track and vertex fitting~\cite{Battaglia:2008nj}. 
Clusters are reconstructed applying a double threshold method on the matrix of pixels around a selected candidate 
cluster seed. Clusters are requested to have a seed pixel with a signal-to-noise ratio, S/N, of at least 7.0 and the 
neighbouring pixels with a S/N in excess of 5.0. Clusters consisting of a single pixel are discarded. The cluster position 
is computed using the centre of gravity of the measured pulse height. The observed average and most probable signal-to-noise 
ratio for clusters associated to reconstructed particle tracks in the pixel doublet is 47.4 and 43.0. The most probable 
value of the signal-to-noise ratio for the seed pixel is 45.1. This can be compared to 45.2$\pm$0.4 predicted by sensor 
simulation~\cite{beam2011-soi}.
The detector planes have been optically surveyed after installation on the beam-line. Their measured positions 
are used as starting point for the offline alignment procedure with particle tracks, performed using the 
{\tt millipede-2} program~\cite{Blobel:2006yh}.

A total of 201000 and 220000 events, having at least one hit per layer in the doublet and only one hit in the singlet, have 
been recorded in the September 2011 data taking with the target installed and 150 and 300~GeV/$c$ beam, respectively. 
An additional $\sim$20000 calibration events have been collected without the target for alignment.

Simulation of interactions and energy release by charge particles in the detectors is performed using the 
{\tt Geant-4} simulation toolkit~\cite{g4} with the {\tt FTFP\_BERT} physics list~\cite{qgsp}, which implements 
high energy inelastic scattering of hadrons by nuclei using the {\tt FRITIOF} model~\cite{Andersson:1992iq}. The yield 
of short- and long-lived particles has been studied using {\tt Pythia 6.125}~\cite{pythia}.
Charge collection and signal generation in the pixels is simulated using a custom {\tt Marlin} processor, 
{\tt PixelSim}~\cite{Battaglia:2007eu}, where the pixel S/N is tuned to the values obtained in data for each chip. 
Simulated events are then analysed through the same reconstruction chain as real data.
 
\subsection{Track Reconstruction}

The efficiency of the SOI sensors is determined by using particle tracks reconstructed on the other two layers and 
extrapolated to the layer under study, in the events taken without the target.

The {\tt Geant-4} + {\tt PixelSim} simulation predicts a sensor efficiency of 0.998$^{+0.002}_{-0.012}$ for the doublet. 
We estimated the efficiency of the sensors in the doublet in an earlier beam test performed in 2010, when the same SOI sensors 
were used both in for doublet and the singlet. We use the number of tracks reconstructed from hits on the second layer of the 
doublet and on the singlet which had an associated hit on the first layer of the doublet and found a value of 
0.98$^{+0.02}_{-0.04}$~\cite{beam2010-soi}.
The single point resolution of the sensors is measured to be (1.12$\pm$0.03)~$\mu$m for the doublet~\cite{beam2010-soi} 
and (1.7~$\pm$0.50)~$\mu$m for the singlet~\cite{beam2011-soi}, where simulation predicts (1.07$\pm$0.04)~$\mu$m and 
(1.63$\pm$0.05)~$\mu$m, respectively. The difference is due to the lower amount of charge collected, and thus of signal-to-noise 
ratio, in the thin sensor. 

Inelastic interaction events are reconstructed from tracks formed with the space points in the two doublet layers.  
We select events with only one reconstructed hit in the singlet layer and at least one accepted hit in each of the 
doublet layers to perform the track reconstruction. This pre-selection removes empty events and most of those with 
more than one primary pion recorded in the same frame. 

\begin{figure}[h!]
\begin{center}
\epsfig{file=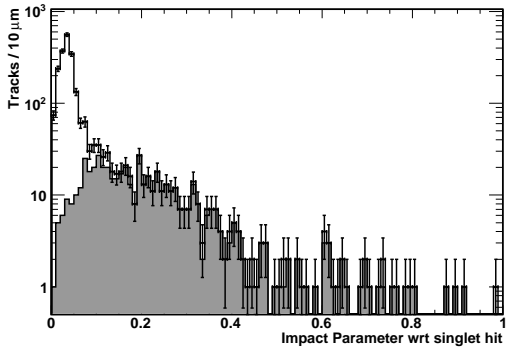,width=7.5cm}
\end{center}
\caption{Impact parameter resolution of tracks reconstructed in the SOI doublet and extrapolated to the singlet hit (in mm).
The distribution of tracks from inelastic events are shown by the filled histogram.}
\label{fig:imp0}
\end{figure}

We reconstruct the trajectory of the incoming pion by projecting the position of the hit reconstructed in the thin singlet. 
Since the r.m.s.\ divergence of the beam is 1.5$\times$10$^{-4}$ rad in the vertical and 7.5$\times$10$^{-4}$ rad in the 
horizontal coordinate, as measured in events taken without the target, the uncertainty on the vertical extrapolation of 
the pion position of impact on the target is $\simeq$3~$\mu$m r.m.s. Secondary candidate tracks are reconstructed in the 
doublet by pairing the hits. These are extrapolated to the singlet and their impact parameter w.r.t.\ the hit recorded on the 
singlet sensor computed. Its distribution is peaked at the value of the extrapolation resolution for primary hadrons, which 
have not interacted in the target, while it has larger values for secondary particles produced in interactions in the target, 
as well as for primary pions experiencing large scattering, as shown in Figure~\ref{fig:imp0}. 
We resolve the combinatorial in the pairing of the doublet hits by choosing the hit pair minimising the impact parameter, 
provided that the longitudinal position of the intercept of the corresponding track with the estimated primary hadron track
is contained in the fiducial region between the singlet layer and the first doublet layer. In simulation, this strategy correctly 
assigns the hit pairing for 
0.96 of the particle tracks emerging from an interaction and contained within the geometrical acceptance of the detector. 
After this first pass, a vertex is reconstructed for events with two or more tracks. Then, candidate tracks are formed 
by pairing the unassociated hits in the doublet and tested for compatibility with the interaction vertex by computing their 
impact parameter w.r.t.\ its position. Candidate tracks with impact parameter smaller than 40~$\mu$m are kept. Again, the 
combinatorial is solved by using the hit pair minimising this impact parameter.

\subsection{Vertex Reconstruction}

The interaction vertex for inelastic interaction events is reconstructed with the candidate tracks fitted in the doublet. 
Several algorithms have been designed to reconstruct the interaction vertex from the trajectories of charged particles. 
They are typically implemented in the context of collider experiments with a homogeneous magnetic field using a helix 
track model. Due to the geometry of the beam test setup and the lack of magnetic field, these implementations have
numerical problems in reconstructing the interaction vertex for our experimental setup. In absence of a magnetic field, 
the track model is a simple straight line with four parameters.  While the track model is simple, a specific implementation of 
the vertex fit procedure is required. Due to the low multiplicity in the interaction events, the vertex position $(x_v,y_v,z_v)$
and the track parameters $p_i = (x'_i,y'_i) = ( (dx/dz)_i, (dy/dz)_i )$ for particle $i$ can be simply derived from the 
minimisation of the sum of $\chi^2_i$ contributions of the trajectory to a global fit. 
The measurement of the trajectory  for particle $i$ on detector $j$ at position $z^j$ is 
given by its Cartesian coordinates $(x^j_i,y^j_i)$ and the corresponding reconstruction errors 
$\sigma_x$ and $\sigma_y$. The $\chi^2_i$ contribution of particle $i$ is then defined as

\begin{equation}
  \chi^2_i = \sum_j 
   \frac{(x^j_i - (x_v + x'_i (z^j-z_v)) )^2}{\sigma^2_x} +
   \frac{(y^j_i - (y_v + y'_i (z^j-z_v)) )^2}{\sigma^2_y}.
\end{equation}
  
Monte Carlo simulation studies show that this approach avoids the numerical shortcomings of other implementations 
of vertex fits. The choice of an adequate track model is essential to reach high efficiency 
and accuracy in the fitting procedure. In this implementation a global fit has been used for simplicity, 
but the straight line track model could also be implemented in more complex vertex reconstruction programs.
\begin{figure}[h!]
\begin{center}
\epsfig{file=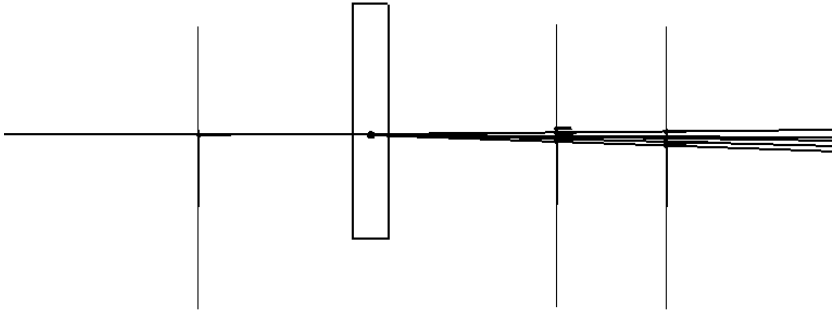,width=11.0cm} 
\end{center}
\caption{A 300~GeV/$c$ $\pi^-$ inelastic interaction event with five charged reconstructed particles. The beam 
comes from the left and the pion position is first determined in the singlet layer, shown on the left. After the interaction 
in the target, in the centre of the picture, charged particle tracks are reconstructed on the two doublet layers shown 
on the right.}
\label{fig:evt}
\end{figure}
The vertex fit is performed iteratively. If the vertex $\chi^2$ probability is smaller than 10$^{-4}$, the track giving 
the largest $\chi^2$ contribution is removed and the vertex recomputed. The hits used for reconstructing these tracks are 
not further used. A reconstructed event is shown in Figure~\ref{fig:evt}. Using fully 
simulated and reconstructed events, we estimate the vertex reconstruction efficiency for inelastic interaction events with 
at least two charged particles within the detector acceptance to be 0.94$\pm$0.05. Further, the reconstructed multiplicity 
of tracks associated to the vertex agrees with the generated charged multiplicity in the detector acceptance within $\pm$0.04.

\section{Results}
 
The number of candidate inelastic interaction events, reconstructed in data by the procedure discussed above, is 1105 
and 2118  for 150 and 300~GeV/$c$ pions, respectively, corresponding to 0.0055$\pm$0.0002 and 0.0096$\pm$0.0006 of the 
events preselected. The observed fractions agree quite well with 
simulation, which predicts 0.0065$\pm$0.0003 and 0.0101$\pm$0.0003, confirming the efficiency indicated by simulation.
For the subsequent analysis we apply additional quality cuts on the reconstructed events requiring the vertex fit 
$\chi^2$ probability to exceed 10$^{-4}$ and the uncertainty on the vertex longitudinal position to be smaller than 1.5~mm.
This yields a total of 1048 selected interaction events at 150~GeV/$c$ and 1862 at 300~GeV/$c$.

\subsection{Tracking Resolution}

\begin{figure}[h!]
\begin{center}
\begin{tabular}{cc}
\epsfig{file=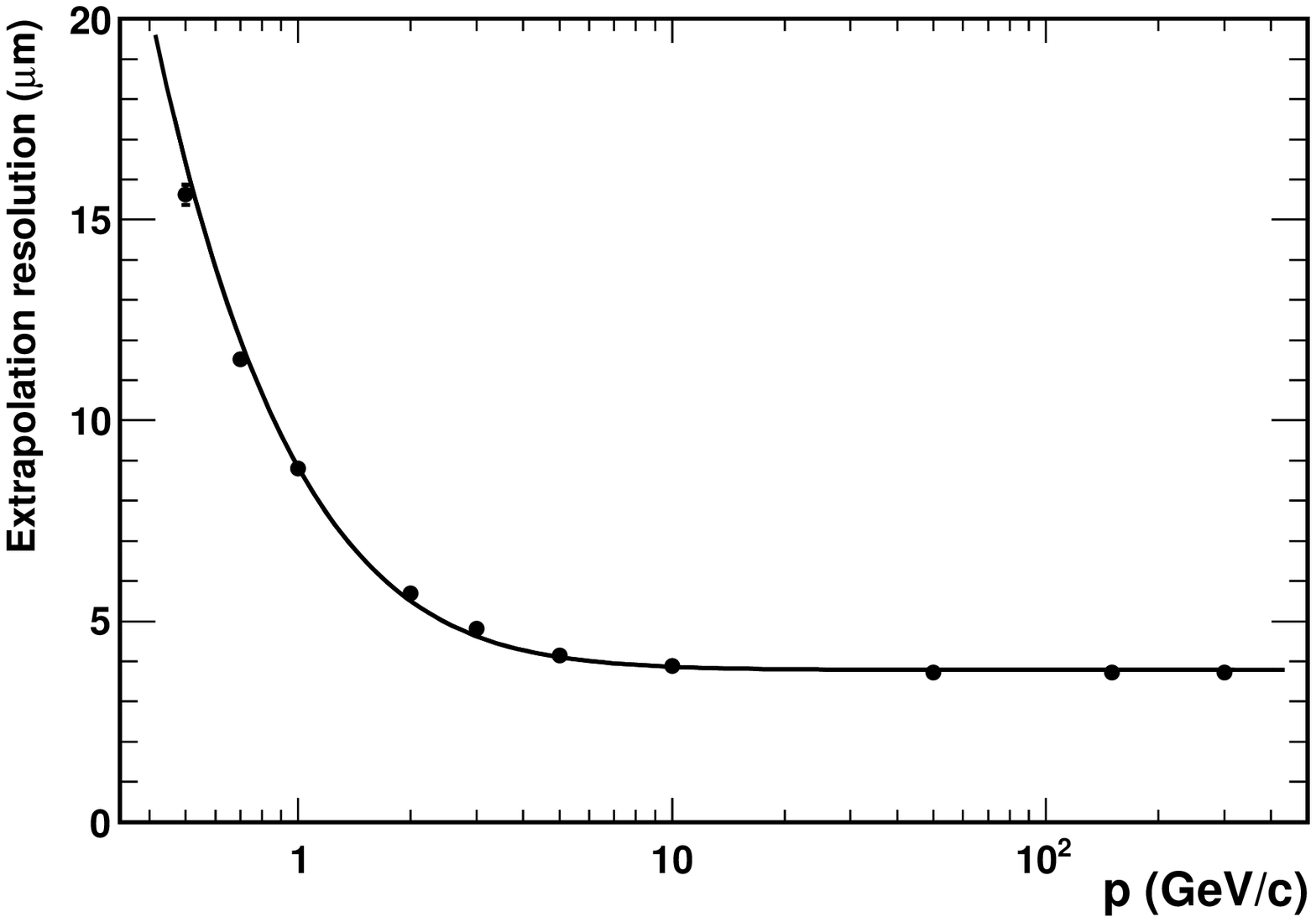,width=6.75cm} &
\epsfig{file=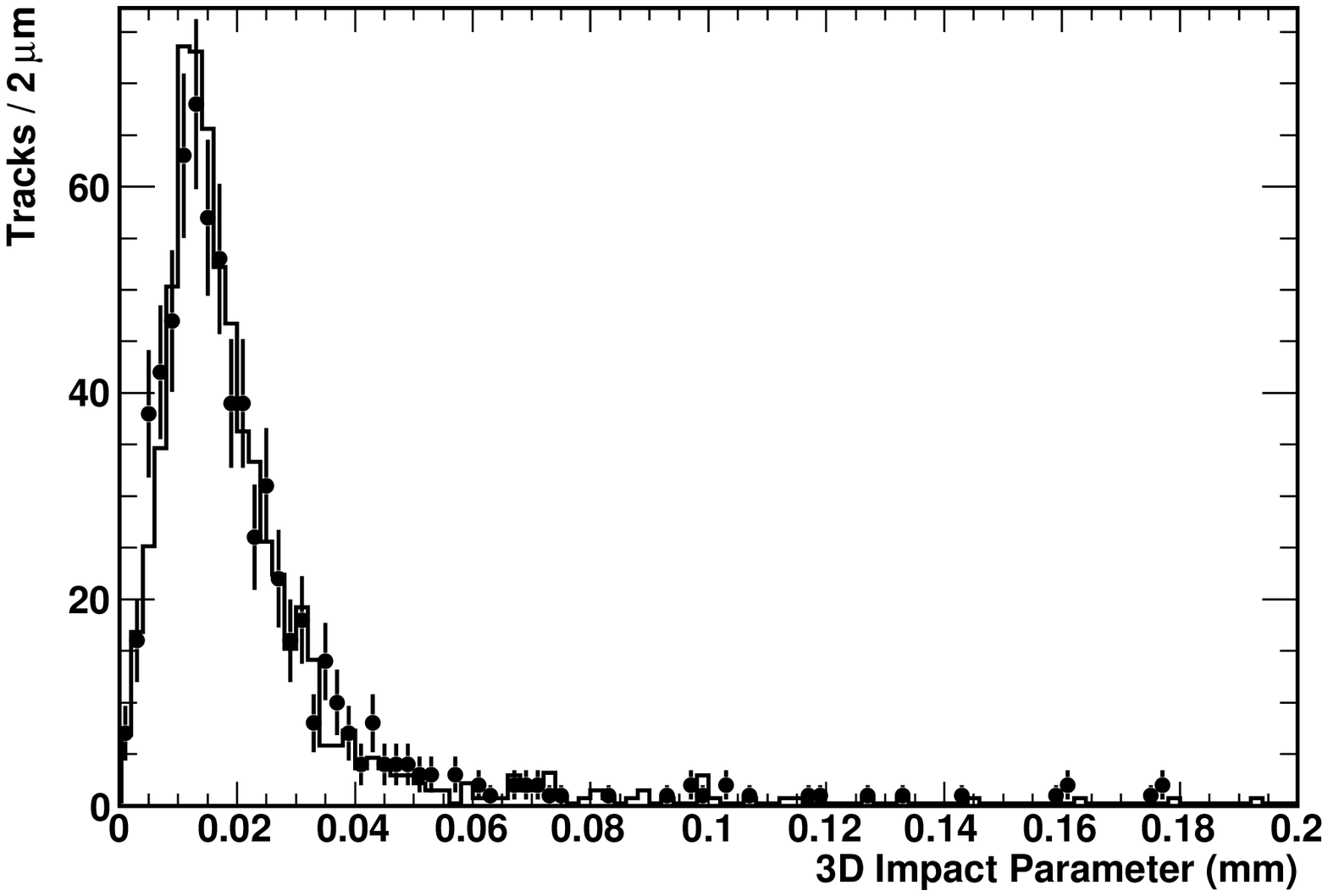,width=6.75cm} \\ 
\end{tabular}
\end{center}
\caption{Track extrapolation resolution on the target using the SOI doublet. Left: the simulation result is shown 
by the dots with error bars while the fitted curve is 3.7~$\mu$m $\oplus \frac{8.0~\mu{\mathrm{m GeV^{-1}}}}{p}$.
Right: impact parameter of tracks w.r.t. the interaction vertex. Tracks are removed from the 
vertex before the impact parameter is computed. Points with error bars represent the data and the line the simulation.}
\label{fig:imp}
\end{figure}

The track extrapolation accuracy at the position of the Cu target, located 15~mm upstream from the first layer of the 
SOI pixel doublet is 3.7~$\mu$m for 300~GeV/$c$ and 8.9~$\mu$m for 1~GeV/$c$ particles according to simulation, as shown 
in Figure~\ref{fig:imp}. 
These values are quite comparable to those expected for the vertex tracker proposed for a linear collider, i.e.\ 
$\simeq$2~$\mu$m $\oplus \frac{10~\mu{\mathrm{m GeV^{-1}}}}{p_t \sin \theta}$, where $\theta$ is the track polar angle, 
for $\sqrt{s}$ = 250 - 500~GeV with innermost pixel layer at $\simeq$15~mm radius, and 
$\simeq$4~$\mu$m $\oplus \frac{21~\mu{\mathrm{m/GeV^{-1}}}}{p_t \sin \theta}$, for $\sqrt{s}$ = 3~TeV  with 
innermost layer at $\simeq$30~mm radius. The impact parameter of tracks w.r.t.\ the reconstructed vertex 
is computed for inelastic interaction events with three or more tracks associated to the vertex. Tracks are iteratively 
removed from the vertex and its position recomputed to avoid the bias due to the contribution of the track under study. 
The distance between the track and the interaction vertex is computed at the position of closest approach. The most probable 
value of the impact parameter resolution is 12.7~$\mu$m, which is the convolution of the track extrapolation resolution, 
the vertex position resolution ($\simeq$9.0~$\mu$m, see section~3.4) and the multiple scattering in the target. 
Data and simulation agree well, as shown in Figure~\ref{fig:imp}.

\subsection{Two-track Separation}

Two-track separation is an important figure of merit for vertex tracking and the optimisation of the pixel sensor geometry. 
In fact, a given target value of the single point resolution can be achieved with different pixel sizes, $P$, depending on 
the S/N ratio and the charge spread on neighbouring pixels. 
In the case of the sensor used here, with $P$ = 13.75~$\mu$m, a seed most probable S/N value of 45 and $\simeq$0.35 of the 
total cluster charge collected on the surrounding pixels, we obtain a 1.1~$\mu$m resolution for isolated particle tracks. 
The same resolution could be achieved also with a pitch $P$ = 25~$\mu$m, for a S/N $\sim$ 70.
On the other hand, two-track separation, which scales $\propto P$, and occupancy, which scales $\propto P^2$, put emphasis 
on small pixel size and limited charge spread to minimise cluster merging. In inelastic interaction events we observe a 
hit density on the first doublet layer of 2.5~hits mm$^{-2}$.
The two-track separation is studied on these events by computing the distance between the two closest reconstructed hits 
on the two layers of the doublet. Due to the large incoming pion momentum, the interaction products are highly boosted in 
the forward direction, closely resembling a highly collimated jet at collider experiments. 
\begin{figure}
\begin{center}
\begin{tabular}{cc}
\epsfig{file=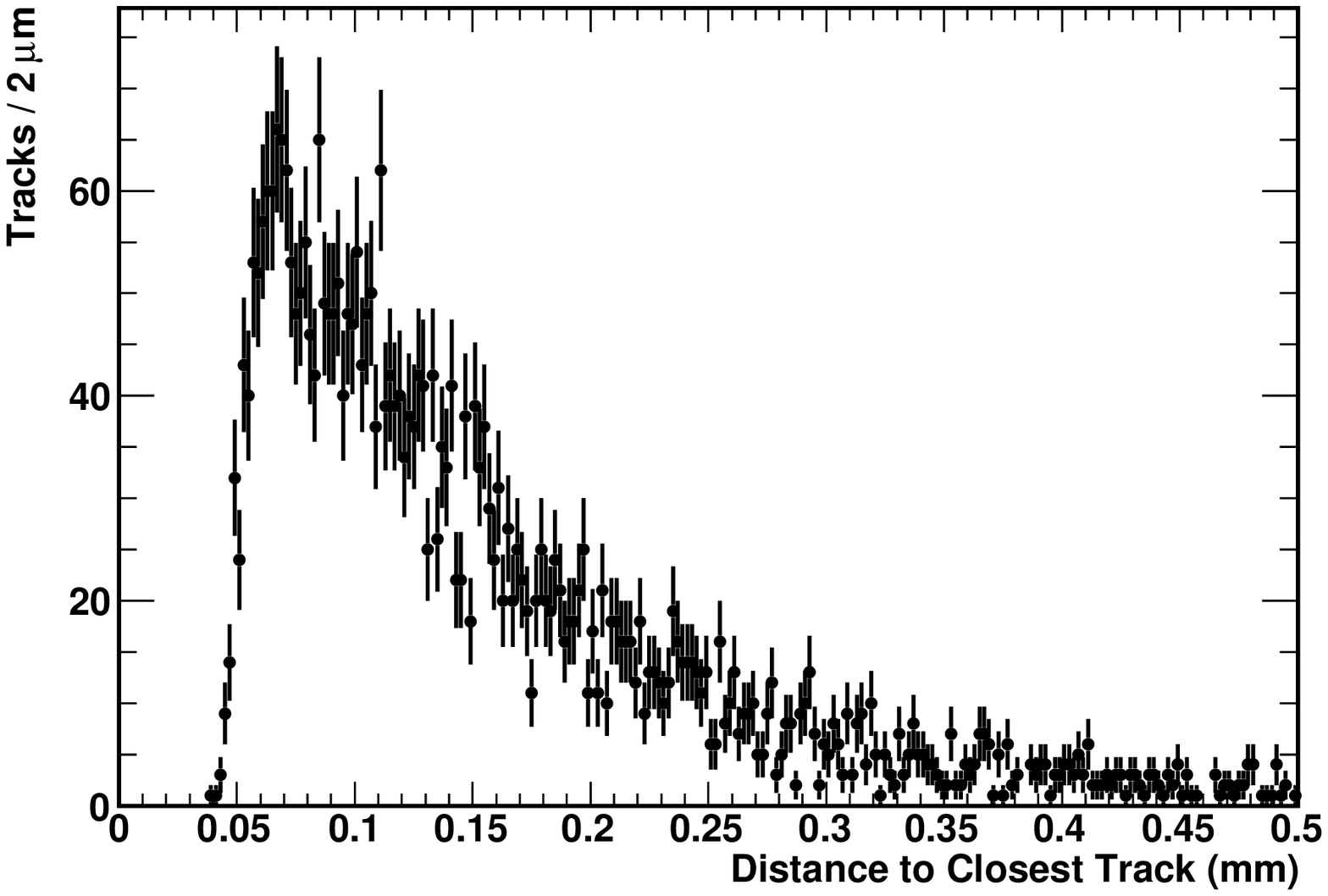,width=6.75cm} &
\epsfig{file=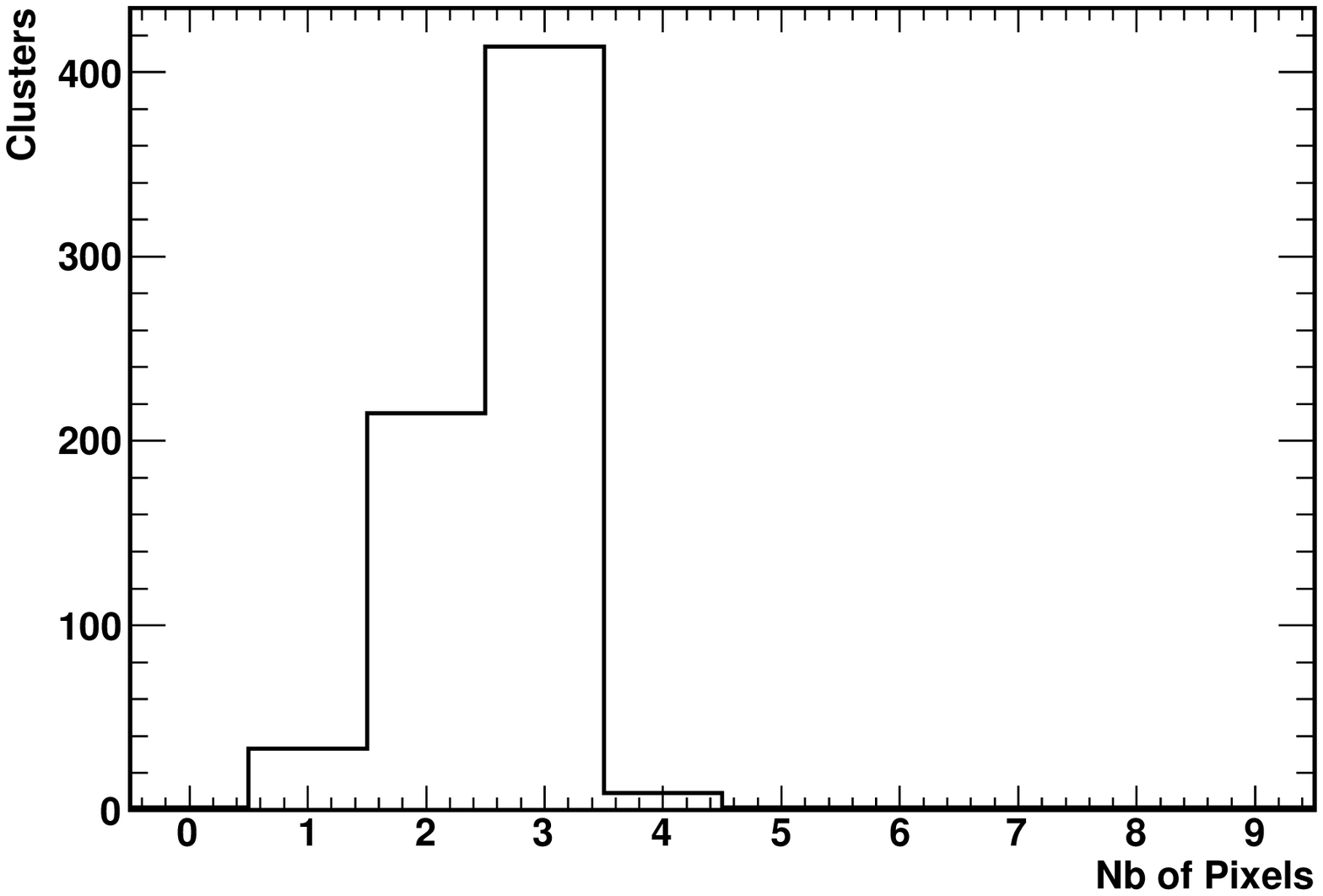,width=6.75cm} \\
\end{tabular}
\end{center}
\caption{Two track separation characterisation. Left: minimum distance between a hit associated to a reconstructed track 
and the closest hit on the first layer of the SOI doublet. Right: number of pixels along rows in clusters associated to 
tracks.}
\label{fig:mindist}
\end{figure}
When they are detected on the first layer of the doublet, the minimum hit distance has a most probable value of 70~$\mu$m. 
In the core of simulated hadronic jets it is $\sim$400~$\mu$m in 
$e^+e^- \rightarrow H^0Z^0 \to b \bar b q \bar q$ events at $\sqrt{s}$ = 500~GeV  and $\sim$240~$\mu$m in
$e^+e^- \rightarrow H^0A^0 \to b \bar b b \bar b$ events at $\sqrt{s}$ = 3~TeV.  
In our data, we observe a distinct cut-off of the distribution at a minimum hit distance value of $\simeq$50~$\mu$m, 
corresponding to the smallest two-track separation resolved in the detector. 
This is comparable to the average size of clusters associated to reconstructed tracks along rows and columns, 
which is 2.67$\pm$0.6 pixels, and to its most probable value, 3 pixels corresponding to 42~$\mu$m 
(see Figure~\ref{fig:mindist}). Below this distance, the charge deposited on the pixels by distinct tracks is merged 
into a single cluster hit.
\begin{figure}[h!]
\begin{center}
\begin{tabular}{cc}
\epsfig{file=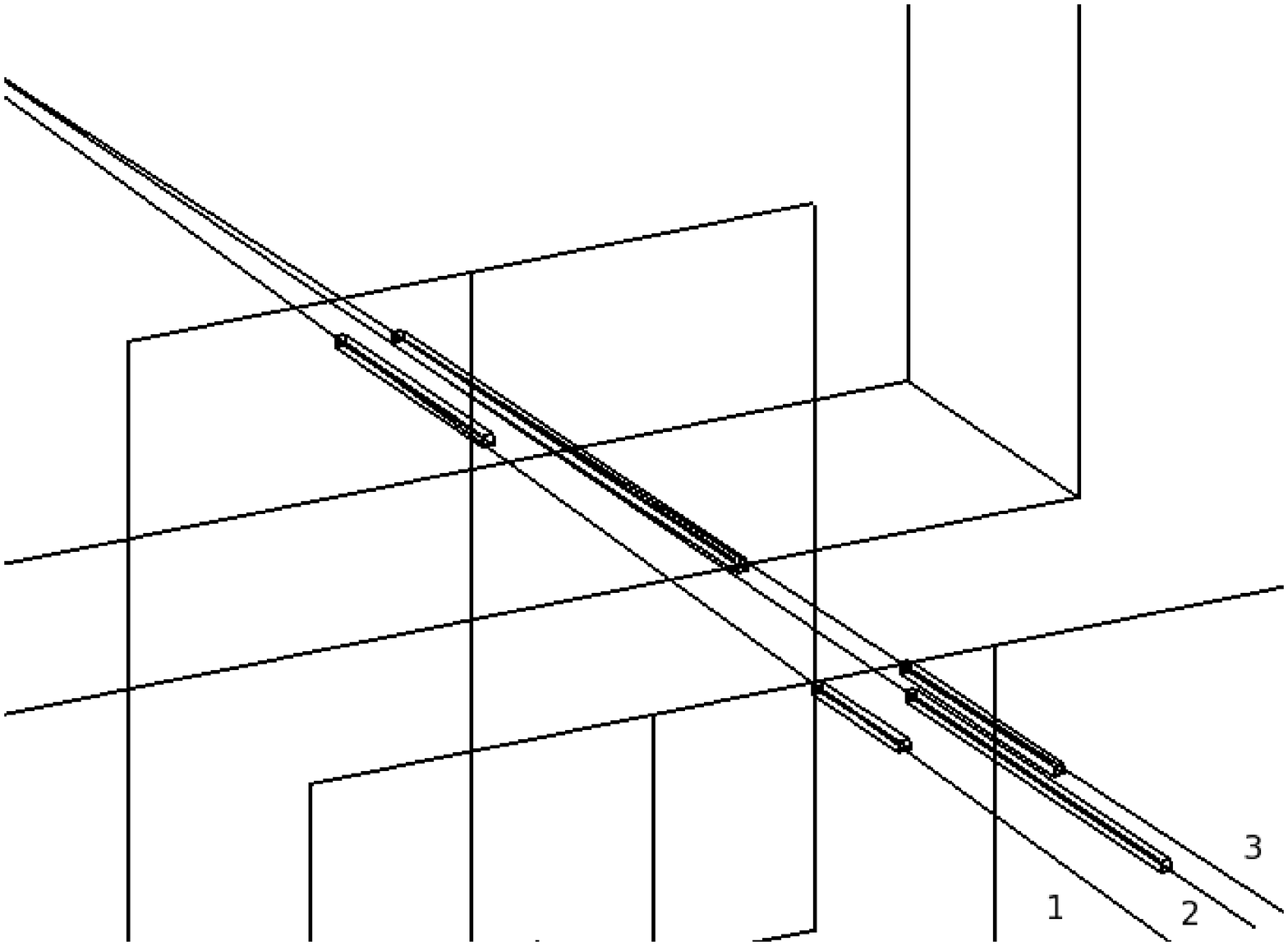,width=6.0cm}
\epsfig{file=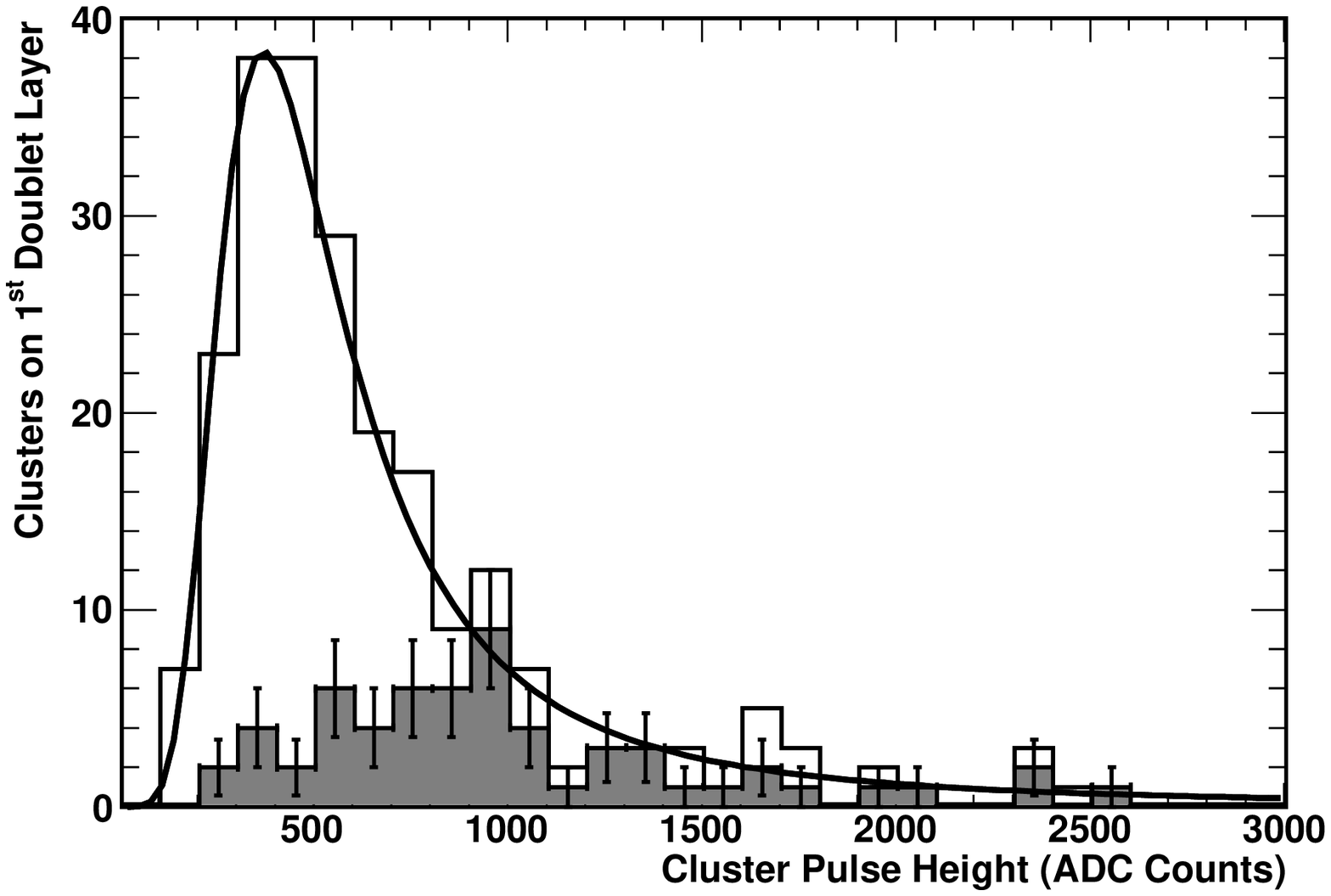,width=7.75cm}
\end{tabular}
\end{center}
\caption{Two track separation and hit merging. Left: display of event with two tracks (labelled 2 and 3) giving a merged 
single hit with cluster pulse height of 1200 ADC counts on the first layer of the SOI doublet. Right: Pulse height 
for clusters on the first layer of the doublet in events with a pseudo-track made with an unassociated hit in the second layer 
(unfilled histogram). The grey filled histogram shows the pulse height distribution for clusters which are within 50~$\mu$m from the 
pseudo-track extrapolation and are already associated to another track. The continuous line shows the fit of a Landau function.}
\label{fig:merge}
\end{figure}
We study the hit merging by selecting events with a reconstructed interaction vertex and at least one hit in the second SOI 
doublet layer which is not associated to a track. In these events, a ``pseudo-track'' is built by fitting a straight line from 
the unassociated hit to the reconstructed vertex. This pseudo-track is extrapolated to the first layer of the doublet and the hit closest to 
the extrapolation point is selected. Figure~\ref{fig:merge} shows one of such events and the cluster pulse height of all these hits 
with highlighted those having a distance from the extrapolated point smaller than 50~$\mu$m, i.e.\ three and half times the pixel pitch, 
and which are already associated to another reconstructed track. This category is characterised by a most probable value of the 
cluster charge of (772$\pm$80)~ADC counts, to be compared to 395$\pm$15 for the distribution of all the selected hits, i.e.\ larger 
by a factor of 1.95$\pm$0.02. 
This shows that tracks impacting on the detector at a distance closer than 3-4 pixels produce merged hit clusters, characterised 
by a most probable value of the collected charge which is approximately double compared to that of isolated clusters. 
This characteristic can be used, in principle, for identifying them. 

It is useful to compare these results to the anticipated track density and two-track distance at high energy lepton 
colliders. These are driven by tracks in collimated hadronic jets and the amount of machine-induced backgrounds, 
mostly incoherent pairs and $\gamma \gamma \to {\mathrm{hadrons}}$ events, integrated in a read-out cycle.   
In the process $e^+e^- \to H^0A^0 \to b \bar b b \bar b$ at $\sqrt{s}$ = 3 TeV, the fraction of particles from a $b$ hadron 
decays reaching the first layer of the Vertex Tracker, at a radius of 30~mm, and within less than 50~$\mu$m from another charged 
particle track is 0.03. The probability to have a charged particle from beam-induced background closer than  50~$\mu$m is 
0.02 assuming a detector integration time of 25~ns, corresponding to 50 bunch crossings. 

\subsection{Interaction Multiplicity}

\begin{figure}[h!]
\begin{center}
\begin{tabular}{cc}
\epsfig{file=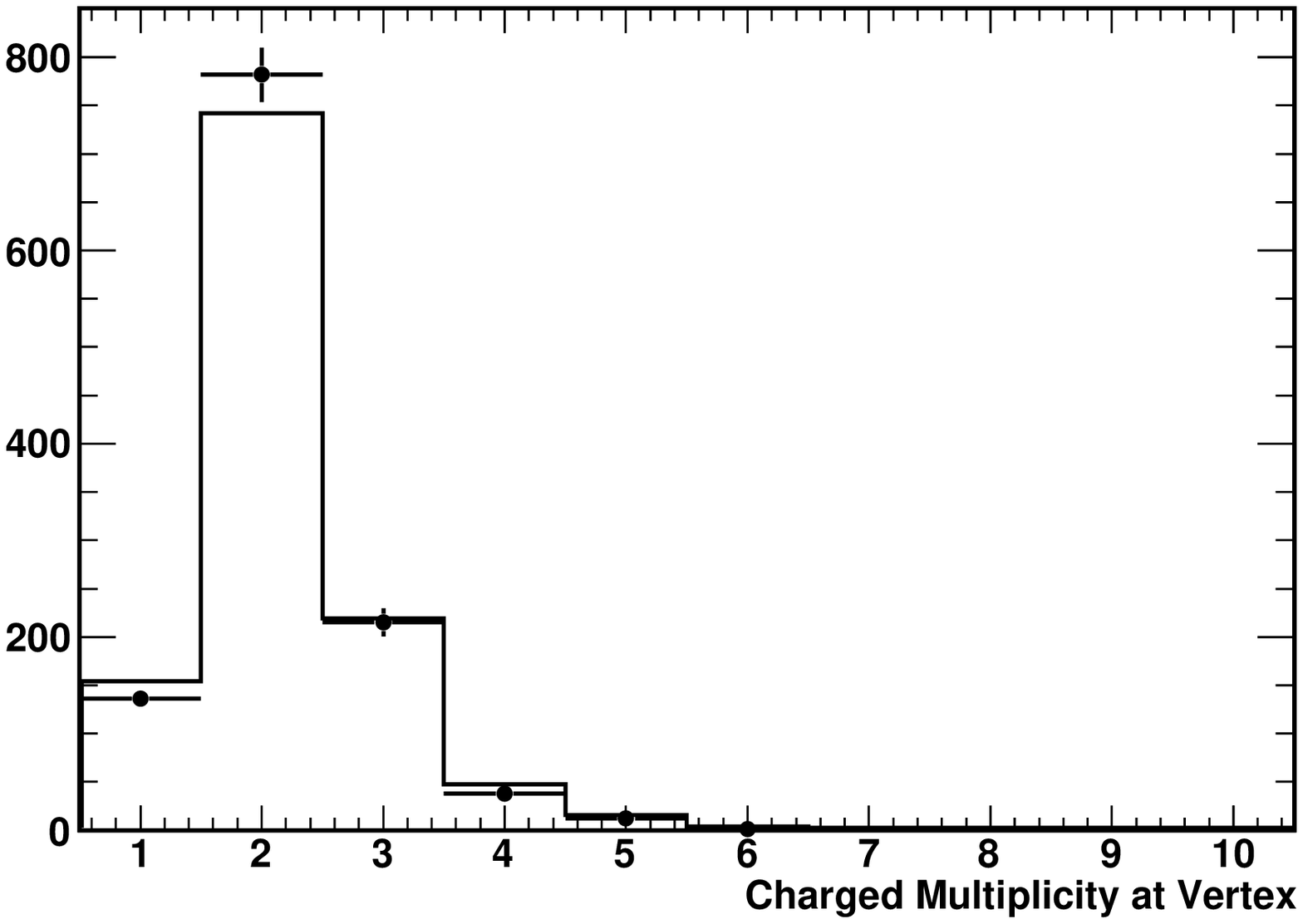,width=6.75cm} &
\epsfig{file=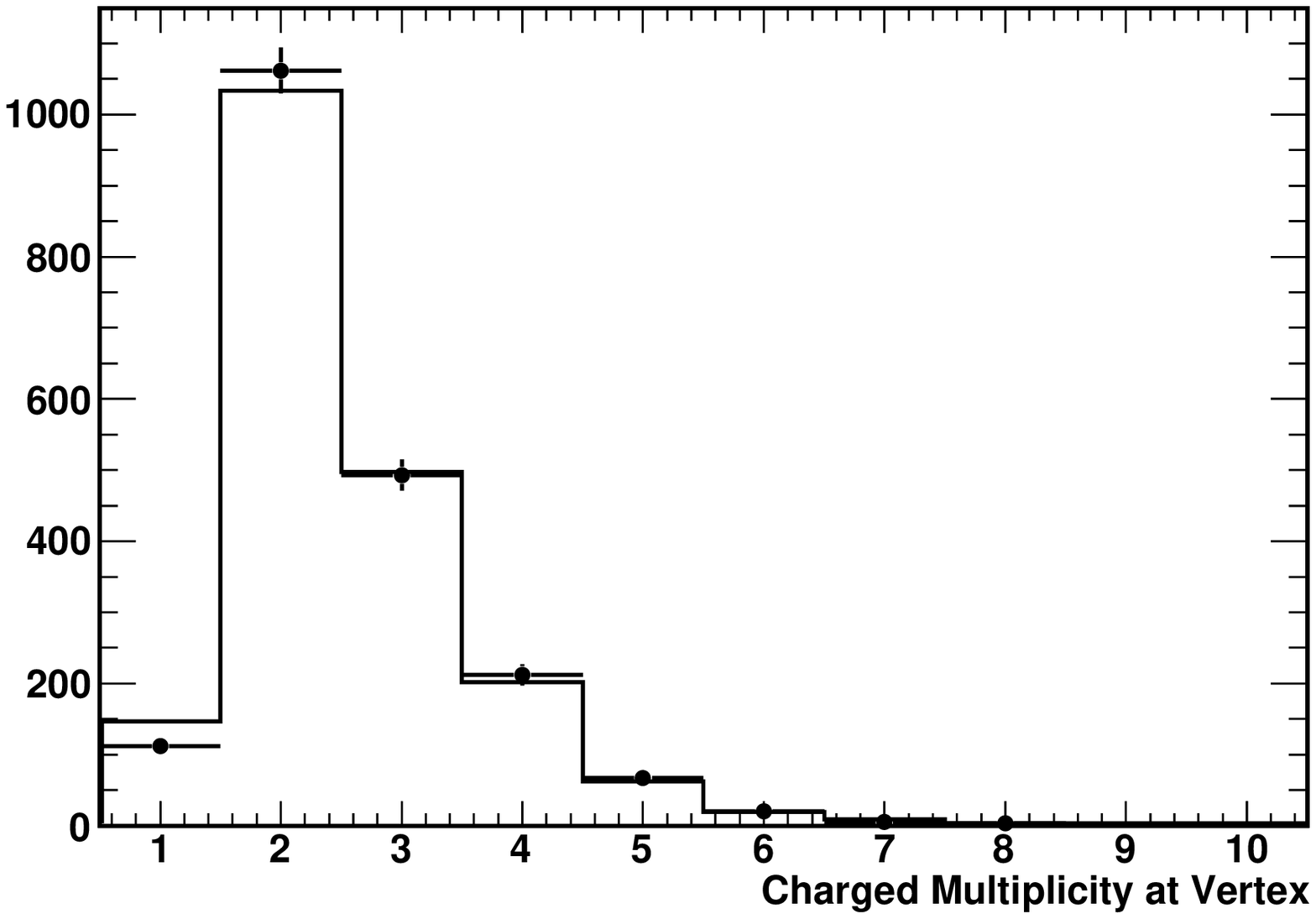,width=6.75cm} \\
\end{tabular}
\end{center}
\caption{Reconstructed charged multiplicity of inelastic interaction events for 150 (left) and 300~GeV/$c$ $\pi^-$s (right). 
Data are shown by points with error bars and simulated and reconstructed events by the continuous line.}
\label{fig:mult}
\end{figure}

The charged multiplicity of inelastic interactions is measured from the number of particles associated to a reconstructed 
vertex, as discussed above. In order to compare with existing data, we include one-prong interaction events. These are 
events having a single reconstructed track in the SOI doublet, which has a longitudinal intercept with the primary pion 
trajectory in the range 16 $< z <$ 23~mm, i.e.\ compatible with the target position, and an impact parameter w.r.t.\ 
the position of the hit in the SOI singlet larger than 100~$\mu$m. We obtain a multiplicity of 2.16$\pm$0.18 and 2.57$\pm$0.22 
for 150 and 300~GeV/$c$ $\pi^-$ beam, respectively (see Figure~\ref{fig:mult}). 
The measured multiplicity scales logarithmically with the beam energy, as expected~\cite{kno}. We compare the multiplicity 
measured in our data as a function of $s$ to a fit to published data scaled to the rapidity range covered by our detector. 
\begin{figure}[h!]
\begin{center}
\epsfig{file=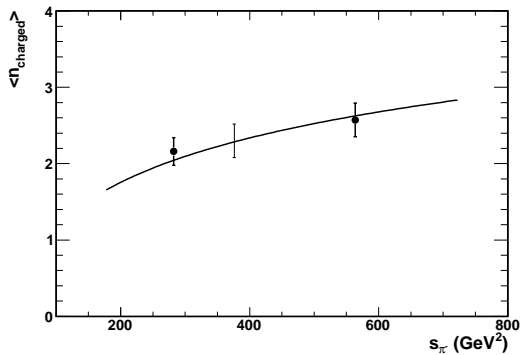,width=7.5cm}
\end{center}
\caption{Average charged multiplicity of inelastic interaction events as a function of $s$. Data are shown by points 
with error bars and the continuous line represents the fit to earlier data rescaled to the same rapidity interval. The error bar 
associated to the line represents the uncertainty of the points used for the fit.}
\label{fig:multS}
\end{figure}
Following~\cite{firestone}, we use a function of the form $<n>$ = $a$ + $b$ ln($s$/$s_0$), where the parameters $a$ and $b$ 
are fitted on results reported for $\pi$~Cu collisions at beam energies from 50 to 200~GeV/$c$~\cite{elias}. We get 
$a$=2.32, $b$=0.84 for $s_0$=392~GeV$^2$. Our results agree with this curve, as shown in Figure~\ref{fig:multS}.

\subsection{Vertex Resolution}

The resolution on the vertex position reconstruction in the plane transverse to the beam is obtained by comparing 
the reconstructed vertex position to that of the incoming pion detected in the thin sensor upstream from the target.
We measure a Gaussian width of (8.9$\pm$0.2)~$\mu$m. By subtracting in quadrature the incident pion extrapolation resolution 
of 3~$\mu$m (see Section 2.1), we obtain (8.4$\pm$0.2)~$\mu$m, which accounts for the vertex resolution, the multiple scattering 
of the incident pion in the target before the interaction and that of the emerging interaction products. The 
estimated average uncertainty from the vertex fit error is 6.2~$\mu$m and the difference between the generated and 
reconstructed position on simulated events has a Gaussian width of 6.5~$\mu$m.

\begin{figure}[h!]
\begin{center}
\epsfig{file=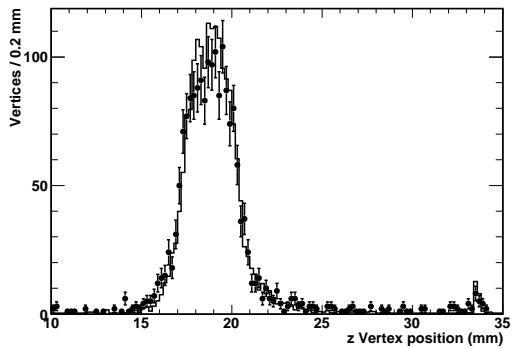,width=7.5cm}
\end{center}
\caption{Distribution of the longitudinal position of fitted vertices in inelastic interactions of 300~GeV/$c$ $\pi^{-}$s. 
The points with error bars show the distribution for data and the continuous line for simulated and reconstructed events.}
\label{fig:vtxz}
\end{figure}

The resolution on the longitudinal vertex position is extracted first from the distribution of the longitudinal position 
of the vertex reconstructed in data and simulation, shown in Figure~\ref{fig:vtxz}. Since the target has sharp edges, 
in absence of resolution effects this distribution is a box function. Resolution effects introduce a smearing of the box 
edges, which can be observed in both data and the simulated and reconstructed interaction events. These vary depending 
on the momentum and multiplicity of the detected particles emerging from the interaction and on the longitudinal position 
of the interaction in the target. 
We perform a multi-parameter $\chi^2$ fit to the distribution of the reconstructed  vertex longitudinal 
position. As a consistency test, we obtain a measurement of the target thickness of (3.15$\pm$0.10)~mm in data and 
(3.07$\pm$0.05)~mm in simulation, where a 3.00~mm thickness was generated. 
The average longitudinal vertex position resolution from the fit at 300~GeV/$c$ is (503$\pm$60)~$\mu$m and (532$\pm$39)~$\mu$m, 
again in data and simulation respectively. It becomes (615$\pm$83)~$\mu$m and (590$\pm$35)~$\mu$m for data and simulation, 
respectively, at 150~GeV/$c$ where the multiplicity is lower and the multiple scattering is more important.

We observe that the distribution of the vertex position away from the target is asymmetric, as we reconstruct more vertices 
downstream from the target than upstream. Since resolution and pattern recognition failures should give spurious vertices 
on either sides of the target position with equal probability, we interpret this as due to secondary vertices of long-lived 
particles produced in the primary interaction and decaying away from the target. Simulation shows a similar effect, mostly 
due to $K_S^0 \to \pi^+ \pi^-$ decays, at a comparable rate as in our data.

In data, the vertex fit error gives a most probable value of (335$\pm$7)~$\mu$m and an average value of (490$\pm$6)~$\mu$m, 
for the longitudinal vertex position resolution at 300~GeV/$c$. The difference in the position of the generated and 
reconstructed vertices in simulated events has a Gaussian width of (290$\pm$10)~$\mu$m and non-Gaussian tails extending 
further, with the contribution of multiple scattering in the target to the Gaussian resolution estimated to be 60~$\mu$m, 
at 300~GeV/$c$.
Given the small angular coverage of our hodoscope and the large boost of the interaction products, the resolution on the 
longitudinal vertex position is driven by the small opening angle of the detected interaction products. In 500~GeV $e^+e^-$ 
collisions, simulated vertices from $B$ mesons with energies in the range 100-150~GeV, re-weighted by the associated 
track multiplicity to match that of the reconstructed vertices in our data, give a resolution of (170$\pm$20)~$\mu$m along 
the heavy meson line of flight~\cite{Battaglia:2008nj}. 
On the other hand, at a multi-TeV collider the typical resolution for high energy jets is larger 
by a factor of two. Decay vertices in $e^+e^- \to H^0A^0 \to b \bar b b \bar b$ events at $\sqrt{s}$ = 3 TeV, where the $B$ 
mesons have energies in the range 200-700~GeV, have been reconstructed using the {\tt ZVTOP} topological vertex 
algorithm~\cite{Bailey:2009ui}. The reconstructed secondary decay vertex has an average charged multiplicity of 2.36 and 
the most probable and average values of the vertex position resolution along the jet axis are (330$\pm$5)~$\mu$m and 
(530$\pm$3)~$\mu$m, respectively. These values are very close to those obtained in the present study.

\section{Conclusions}

Tracking and vertexing performance of a small beam hodoscope made of SOI pixel sensors have been determined in a beam 
test with 150 and 300~GeV/$c$ $\pi^-$s at the CERN SPS. The charged multiplicity of inelastic $\pi$-Cu interactions has 
been determined to be  2.16$\pm$0.18 at 150~GeV/$c$ and 2.57$\pm$0.22 at 300~GeV/$c$, which agrees with earlier data and 
logarithmic scaling with the beam energy. The two-track separation of the sensor is $\simeq$50~$\mu$m, i.e.\ four pixels, 
which agrees with the observed pixel multiplicity along columns and rows in reconstructed clusters. Merged hits can be 
identified by the large collected charge in the cluster. The vertex resolution is estimated to be 8.4~$\mu$m transverse 
and $\simeq$350-500~$\mu$m longitudinal. The results obtained are representative of the experimental conditions and 
reconstruction performance expected in highly boosted hadronic jets at TeV-class lepton colliders and demonstrate that 
pixels sensors with $\sim$15~$\mu$m pitch on high-resistivity Si are extremely well suited for these applications.

\section*{Acknowledgements}

This work was supported by the Director, Office of Science, of the 
U.S. Department of Energy under Contract No.DE-AC02-05CH11231 and 
by INFN, Italy. We are grateful to Y.~Arai for his effective 
collaboration in the SOIPIX activities, D.~Bisello for 
his support and P.~Giubilato for contributions to the data 
acquisition system. We are indebted to I.~Efthymiopoulos and  M.~Jeckel 
for support on the SPS beam-line and to A.~Behrens and E.~Lacroix for 
performing the detector alignment. We also thank the CERN IT department 
for computing support.


\begin{thebibliography}{99}
%
\bibitem{Hansen:2003sb}
  S.~Xella~Hansen  [LCFI Collaboration],
  %``Flavour tagging at the future linear collider,''
  Nucl.\ Instrum.\ and Meth.\  A {\bf 501} (2003) 106.
  %%CITATION = NUIMA,A501,106;%%

\bibitem{Battaglia:2010ce}
  M.~Battaglia,
  Nucl.\ Instrum.\ and Meth.\  A {\bf 650} (2011) 55.

\bibitem{ATLAS}
  G.~Aad {\it et al.} [ATLAS Collaboration], Phys.\ Lett.\ B {\bf 710} (2012) 49.

\bibitem{cms}
  S.~Chatrchyan {\it et al.} [CMS Collaboration], Phys.\ Lett.\ B {\bf 710} (2012) 26.

\bibitem{tevnph}
  The TEVNPH Working Group for the CDF and D0 Collaborations, arXiv:1203.3774 [hep-ex].

\bibitem{Dolezal:2010zz}
  Z.~Dolezal [Belle~II Collaboration],
  %``Super KEKB and Belle II: status of the KEK SuperB factory,''
  PoS HQL {\bf 2010} (2010) 079.
  %%CITATION = POSCI,HQL2010,079;%%

\bibitem{Bona:2007qt}
  M.~Bona {\it et al.}  [SuperB Collaboration],
  %``SuperB: A High-Luminosity Asymmetric e+ e- Super Flavor Factory. Conceptual Design Report,''
  Report INFN/AE-07/2, arXiv:0709.0451 [hep-ex].
  %%CITATION = ARXIV:0709.0451;%%

\bibitem{Forti:2011zz}
  F.~Forti {\it et al.},
  %``The SuperB silicon vertex tracker,''
  Nucl.\ Instrum.\ and Meth.\ A {\bf 636} (2011) S168.
  %%CITATION = NUIMA,A636,S168;%%

\bibitem{Besson:2006dd}
  A.~Besson {\it et al.},
  %``A vertex detector for the International Linear Collider based on CMOS sensors,''
  Nucl.\ Instrum.\ and Meth.\ A {\bf 568} (2006) 233.
  %%CITATION = NUIMA,A568,233;%%

\bibitem{Marinas:2011zz}
  C.~Marinas and M.~Vos,
  %``The Belle-II DEPFET pixel detector: A step forward in vertexing in the superKEKB flavour factory,''
  Nucl.\ Instrum.\ and Meth.\ A {\bf 650} (2011) 59.
  %%CITATION = NUIMA,A650,59;%%

\bibitem{Rizzo:2011zz}
  G.~Rizzo {\it et al.},
  %``Thin pixel development for the SuperB silicon vertex tracker,''
  Nucl.\ Instrum.\ and Meth.\ A {\bf 650} (2011) 169.
  %%CITATION = NUIMA,A650,169;%%

\bibitem{Dorokhov:2011zz}
  A.~Dorokhov {\it et al.},
  %``High resistivity CMOS pixel sensors and their application to the STAR PXL detector,''
  Nucl.\ Instrum.\ and Meth.\ A {\bf 650} (2011) 174.
  %%CITATION = NUIMA,A650,174;%%

\bibitem{Deveaux:2011zz}
  M.~Deveaux {\it et al.},
  %``Radiation tolerance of a column parallel CMOS sensor with high resistivity epitaxial layer,''
  JINST {\bf 6} (2011) C02004.
  %%CITATION = JINST,6,C02004;%%

\bibitem{Marczewski:2004zz}
  J.~Marczewski {\it et al.},
  %``SOI active pixel detectors of ionizing radiation-technology and design development,''
  IEEE Trans.\ Nucl.\ Sci.\  {\bf 51} (2004) 1025.
  %%CITATION = IETNA,51,1025;%% 
\bibitem{soi2}
  Y.~Arai {\it et al.},
  Nucl.\ Instrum.\ and Meth.\  A {\bf 623} (2010) 197.

\bibitem{beam2010-soi}
  M.~Battaglia {\it et al.},   
  Nucl.\ Instrum.\ and Meth.\  A, {\bf 654} (2011) 258.

\bibitem{beam2011-soi}
  M.~Battaglia {\it et al.},  Nucl.\ Instrum.\ and Meth.\  A {\bf 676} (2012) 51. 

\bibitem{xray-soi}
  M.~Battaglia {\it et al.},  Nucl.\ Instrum.\ and Meth.\  A, {\bf 674} (2012) 50. 

\bibitem{Battaglia:2009daq}
  M.~Battaglia {\it et al.},
  Nucl.\ Instrum.\ and Meth.\  A {\bf 611} (2009) 105.

\bibitem{Brun:1997pa}
  R.~Brun and F.~Rademakers,
  %``ROOT: An object oriented data analysis framework,''
  Nucl.\ Instrum.\ and Meth.\  A {\bf 389} (1997) 81.
  %%CITATION = NUIMA,A389,81;%%

\bibitem{Gaede:2005zz}
  F.~Gaede {\it et al.},
  %``LCIO persistency and data model for LC simulation and reconstruction,''
  %\href{http://www.slac.stanford.edu/spires/find/hep/www?irn=7872623}{SPIRES entry}
  in Proc. of {\it  Interlaken 2004, Computing in high energy physics and nuclear physics}, 
  Report CERN 2005-002, 471.

\bibitem{Gaede:2006pj}
  F.~Gaede,
  %``Marlin and LCCD: Software tools for the ILC,''
  Nucl.\ Instrum.\ and Meth.\ A {\bf 559} (2006) 177.
  %%CITATION = NUIMA,A559,177;%%

\bibitem{Battaglia:2008nj}
  M.~Battaglia {\it et al.},
  %``Tracking and Vertexing with a Thin CMOS Pixel Beam Telescope,''
  Nucl.\ Instrum.\ and Meth.\  A {\bf 593} (2008) 292.
  %%CITATION = NUIMA,A593,292;%%

\bibitem{Blobel:2006yh}
  V.~Blobel,
  %``Software alignment for tracking detectors,''
  Nucl.\ Instrum.\ and Meth.\ A {\bf 566} (2006) 5.
  %%CITATION = NUIMA,A566,5;%%

\bibitem{g4}
  S.~Agostinelli {\it et al.},  
  Nucl.\ Instrum.\ and Meth.\ A {\bf 506} (2003) 250.

\bibitem{qgsp}
  J.~Apostolakis {\it et al.}, J.\ Phys.\ Conf.\ Series {\bf 160} (2009) 012073.

\bibitem{Andersson:1992iq}
  B.~Andersson, G.~Gustafson and H.~Pi,
  %``The FRITIOF model for very high-energy hadronic collisions,''
  Z.\ Phys.\ C {\bf 57} (1993) 485.
  %%CITATION = ZEPYA,C57,485;%%

\bibitem{pythia}
  T. Sj\"ostrand {\it et al.}, Comp.\ Phys.\ Commun.\ {\bf 135} (2001) 238. 

\bibitem{Battaglia:2007eu}
  M.~Battaglia,  
  Nucl.\ Instrum.\ and Meth.\ A {\bf 572} (2007) 274.

\bibitem{kno}
  Z.~Koba, H.B.~Nielsen and P.~Olesen, Nucl.\ Phys.\ B {\bf 40} (1972) 317.

\bibitem{firestone}
  A.~Firestone {\it et al.}, Phys.\ Rev.\ D {\bf 14} (1976) 2902.

\bibitem{elias}
  J.E. Elias {\it et al.}, Phys.\ Rev.\ D {\bf 22} (1980) 13.

\bibitem{Bailey:2009ui}
  D.~Bailey {\it et al.}  [LCFI Collaboration],
  %``The LCFIVertex package: vertexing, flavour tagging and vertex charge reconstruction with an ILC vertex detector,''
  Nucl.\ Instrum.\ and Meth.\ A {\bf 610} (2009) 573.

\end{thebibliography}
\end{document}